\documentclass[epj]{svjour}
\usepackage{graphics}
\begin{document}
\title{Modified Fragmentation Function in Heavy Ion Collisions at 
RHIC via Direct $\gamma$-Jet Measurements}
\author{ A. M. Hamed\thanks{\email{ahamed@tamu.edu}}
for the STAR Collaboration
}                     
\institute{Texas A$\&$M University, Physics Department, Cyclotron Institute, College Station, TX 77843, USA}
\date{Received: date / Revised version: date}
%
\onecolumn
\abstract{The presented results are the  
first measurements at RHIC for direct $\gamma$-charged hadron azimuthal correlations in heavy ion collisions. We use these correlations
to study the color charge density of the medium through the medium-induced modification of high-p$_T$ parton fragmentation.
Azimuthal correlations of direct photons at high transverse energy (8 $<$ p$_T$ $<$ 16 GeV) 
with away-side charged hadrons of transverse momentum (3 $<$ p$_T$ $<$ 6 GeV/c) have been measured over a broad range of centrality for 
$Au+Au$ collisions and $p+p$ collisions at $\sqrt{s_{NN}}$ = 200 GeV in the STAR experiment. 
A transverse shower shape analysis in the STAR Barrel Electromagnetic Calorimeter Shower Maximum Detector is used to discriminate between the 
direct photons and photons from the decays of high p$_T$ $\pi^{0}$. 
The per-trigger away-side yield of direct $\gamma$ is smaller than from $\pi^{0}$ trigger at the same centrality class. 
Within the current uncertainty the I$_{CP}$ of direct $\gamma$ and $\pi^{0}$ are similar.
} 
\titlerunning{Direct $\gamma$-Jet Measurements}
\authorrunning{A. M. Hamed}
\maketitle
\section{Introduction}
\label{intro}
Despite the lack of understanding of the 
early time dynamics, including the modifications of the accelerated nuclei even before the collisions, various complementary 
measurements from RHIC data 
have revealed the formation of a strongly coupled medium in contrast to the expected weakly coupled medium [1,2,3,4,5].
Determining the characteristics of the formed medium is the main goal for the current RHIC heavy ion program. One of 
the most important of such characteristics is the 
medium color charge density, which might lead to understand the medium dynamics. Actually, 
the color structure of the medium can be probed by its 
effect on the propagation of a fast parton [6]. The characterization of the 
resulting medium-induced modification of high-p$_T$ parton fragmentation 
has become one of the most 
active areas of research stimulated by RHIC data at the
experimental and theoretical levels.

The first three years of RHIC have produced one of the most celebrated results at RHIC which is the suppression of hadrons in central $Au+Au$ collisions, 
compared to $p+p$ collisions [7] and to cold nuclear matter [8] at high p$_{T}$. This suppression is well described by very different
pQCD-based energy loss models in the light flavor sector [9]. However, 
the observation of a similar level of suppression for the heavy flavor [10], and the unexpected smaller suppression of baryons compared to mesons  
at intermediate to high p$_{T}$ [11], cannot be easily reconciled in these pQCD-based energy loss models that incorporate 
gluon radiation as the main source of in-medium energy loss. In fact the applicability of pQCD in describing the parton-matter 
interaction has been increasingly challenged by the strongly coupled nature of the produced matter at RHIC. As a result there is no single commonly accepted calculation of 
the underlying physics to describe in-medium energy loss for different quark generations as well as for the gluon.
 
Experimental observables based on single-particle spectra are not sensitive enough to discriminate 
between the different energy loss mechanisms [12], indicating the need for additional experimental observables in order to better
constrain the energy loss mechanism. The
di-hadron azimuthal correlation measurement is expected to provide a slightly better way in constraining the energy loss since the initial jet energy is not 
accessible in single-particle spectra but is somewhat accessible in the di-hadron measurements.     
More model dependent detailed studies [13] 
have shown that the azimuthal correlation measurements of
di-hadrons are more sensitive than single particle spectra but both have diminished sensitivity at high gluon density due to the
geometrical bias. The $\gamma$-hadron azimuthal correlations measurement provides a powerful way to 
quantify the energy loss through the dominant
process QCD Compton-like scattering since the photon transverse momentum balances the parton initial energy [14]. 
In particular the $\gamma$-hadron azimuthal correlations provide a unique way to quantify the energy loss 
dependence on the initial parton energy and possibly the color-factor. 

\section{Data Analysis}
\label{data analysis}
Using a level-2 high-$p_T$ tower trigger to tag $\gamma$-jet events, in 2007 the STAR experiment collected an integrated
luminosity of 535 $\mu {b}^{-1}$ of $Au+Au$ collisions at $\sqrt{s_{NN}}$ = 200 GeV. The level-2 trigger algorithm was implemented in the Barrel Electromagnetic
Calorimeter (BEMC) and optimized based on the information of the direct $\gamma/\pi^0$ ratio in $Au+Au$ collisions [15], the
$\pi^0$ decay kinematics, and the electromagnetic shower profile characteristics. The BEMC has a full azimuthal coverage and
pseudorapidity coverage $\mid\eta\mid$ $\leq$ 1.0. As a reference measurement we use $p+p$ data at $\sqrt{s_{NN}}$ = 200 GeV
taken in 2006 with integrated luminosity of 11 $\mathrm{pb}^{-1}$. The Time Projection Chamber (TPC) was used
to detect charged particle tracks and measure their momenta. The charged track quality cuts are similar
to previous STAR analyses [16]. For this analysis, events with at least one cluster with $E_T >$ 8~GeV were selected. 
To ensure the purity of the photon triggered sample trigger towers were rejected if a track with $p >$ 3~GeV/$c$ points to it.
 
A crucial step of the analysis is to discriminate between showers of direct $\gamma$ and two close $\gamma$'s from a
high-p$_{T}$ $\pi^{0}$ symmetric decay. At p$_T$ $\sim$ 8 GeV/c the angular separation between the two photons 
resulting from a $\pi^{0}$ symmetric decay (both decays photons have similar energy, smallest opening angle) at the BEMC face is typically smaller than the tower size ($\Delta\eta=0.05,\Delta\phi=0.05$); but 
a $\pi^{0}$ shower is generally broader than a single $\gamma$ shower. The Barrel Shower Maximum Detector (BSMD), 
which resides at $\sim$ 5X$_{0}$ inside the calorimeter towers, is well-suited for 
$(2\gamma$)/$(1\gamma)$ separation up to p$_T$ $\sim$ 26 GeV/c due to its fine segmentation ($\Delta\eta\approx 0.007,\Delta\phi\approx 0.007$). 
In this analysis the $\pi^{0}$/$\gamma$ discrimination was carried out by making cuts on the 
shower shape as measured by the BSMD, where the $\pi^{0}$ identification 
cut is adjusted in order to obtain very pure sample of $\pi^{0}$ and a sample rich in direct $\gamma$ ($\gamma_{rich}$). 
The discrimination cuts are varied to determine the systematic uncertainties. To determine the combinatorial background level the relative azimuthal angular distribution of the associated particles with respect to the trigger particle 
per trigger is fitted with a two guassian peaks and a straight line.  
The near- and away-side yields, Y$^{n}$ and Y$^{a}$, of associated particles per trigger are extracted by 
integrating the $\rm 1/N_{trig} \rm dN/{\rm d}(\Delta\phi)$  
distributions in $\mid\Delta\phi\mid$ $\leq$  0.63 and $\mid\Delta\phi -\pi\mid$  $\leq$  0.63 respectively. 
The yield is corrected for the tracking efficiency of associated charged particles as a function of multiplicity. 

The shower shape cuts used to select a sample of direct-photon-``rich" triggers reject most of the $\pi^{0}$'s, but do not reject 
photons from highly asymmetric $\pi^{0}$ decays, $\eta$'s, and fragmentation photons. 
All of these sources of background are removed as follows (Eq 1), but only within the systematic
uncertainty on the assumption that their correlations are similar to those for $\pi^{0}$'s.
Assuming zero near-side yield for direct photon triggers and a very pure sample of $\pi^{0}$, 
the away-side yield of hadrons correlated with the direct photon is extracted as
\begin{eqnarray}
Y_{\gamma_{direct}+h}=\frac{Y^{a}_{\gamma_{rich}+h}-R Y^{a}_{\pi^{0}+h}}{1-R},\label{eq2}
\hspace{0.5cm}                            
   \hspace{0.5cm}   R=\frac{Y^{n}_{\gamma_{rich}+h}}{Y^{n}_{\pi^{0}+h}}.
\end{eqnarray}
Where Y$^{a(n)}_{\gamma_{rich}+h}$ and Y$^{a(n)}_{\pi^{0}+h}$ are the away (near)-side yields of associated particles 
per $\gamma_{rich}$ and $\pi^{0}$ triggers
respectively, so that R is the fraction of $\gamma_{rich}$ triggers that are actually from $\pi^{0}$, $\eta$, and fragmentation
photons.
\section{Results}
\label{results}

\begin{figure}
\begin{center}
\begin{tabular}{cc}
   \resizebox{90mm}{153pt}{\includegraphics{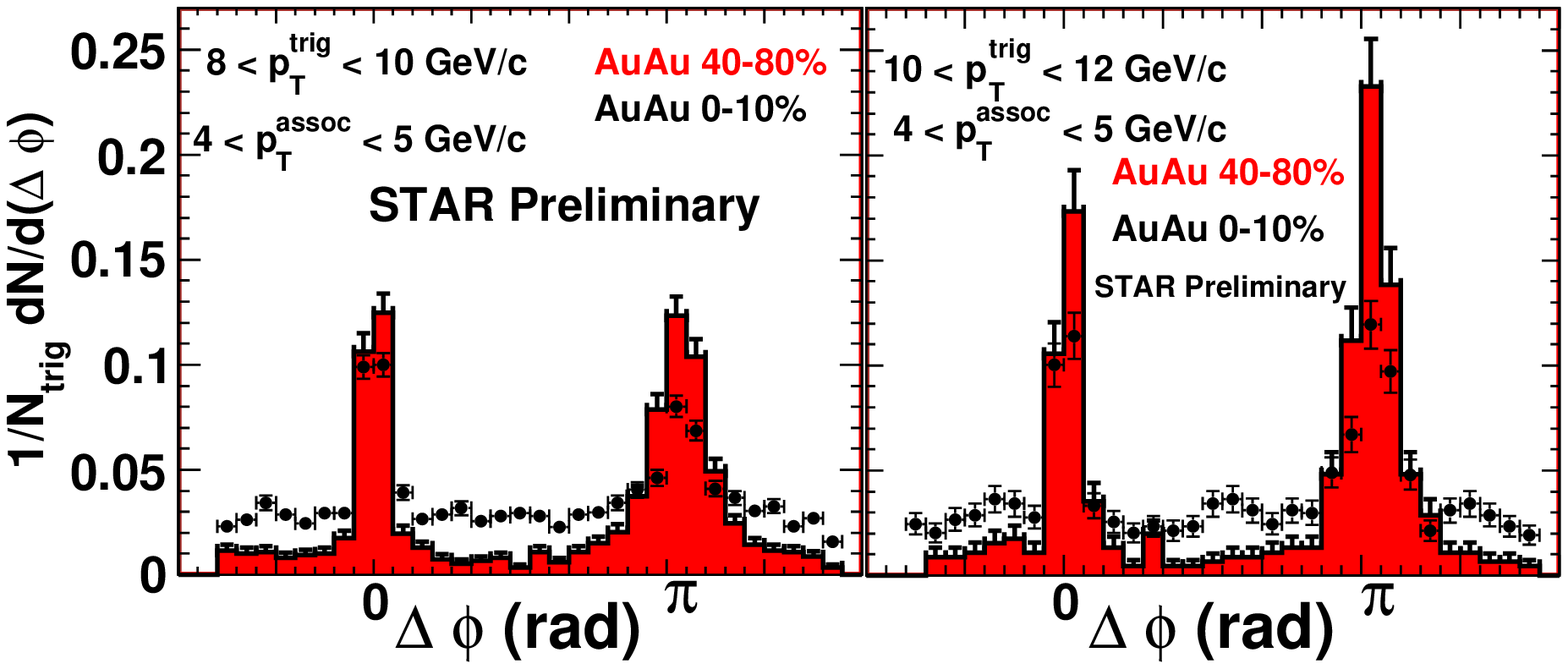}} 
    \resizebox{90mm}{160pt}{\includegraphics{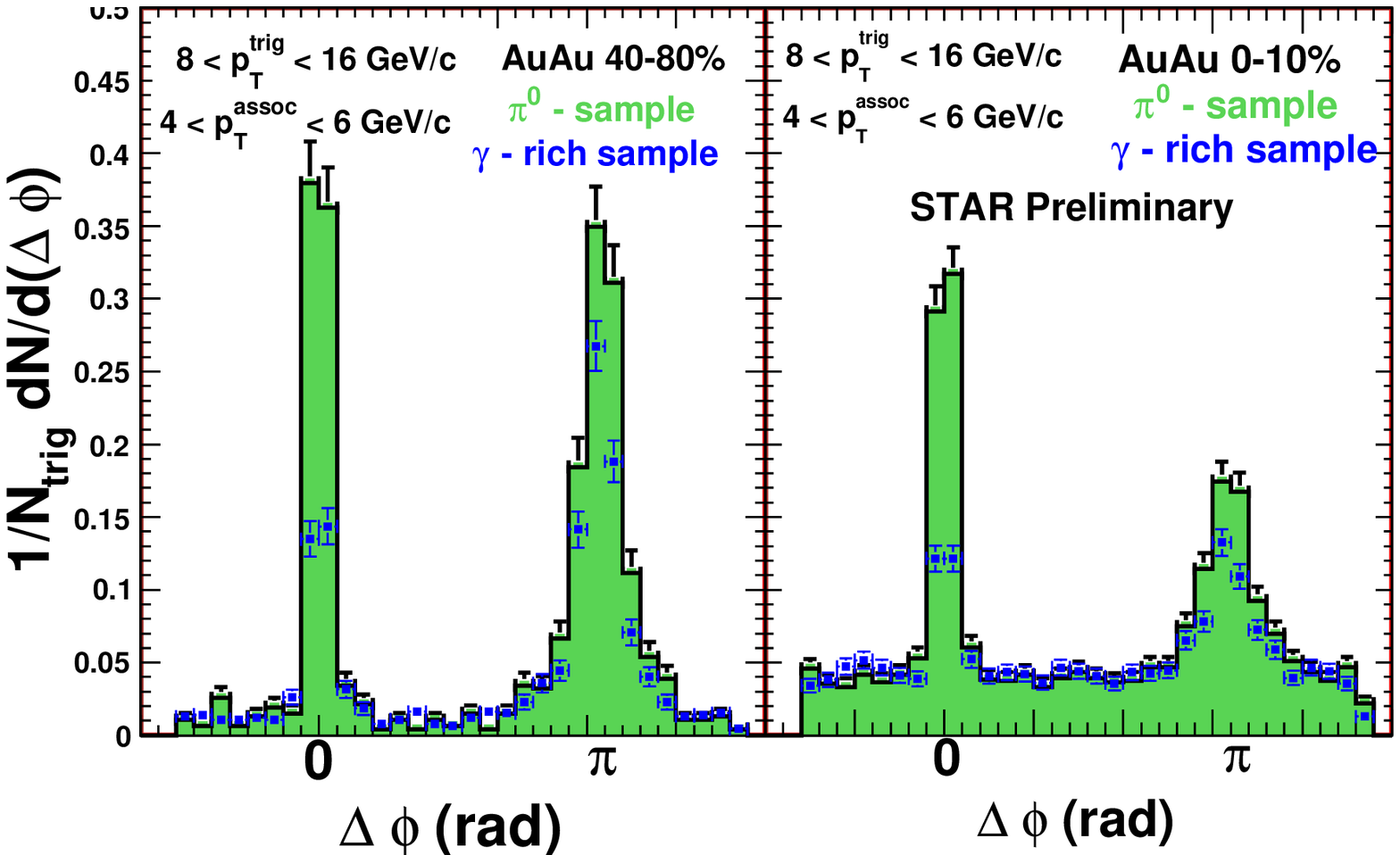}} \\
        \end{tabular}
    \caption{Left: Azimuthal correlation histograms of high p$_{T}^{trig}$ inclusive photons 
    with associated hadrons for 40-80$\%$ and 0-10$\%$ $Au+Au$
    collisions. Right: Azimuthal correlation histograms of high p$_{T}^{trig}$ $\gamma$-rich sample and $\pi^{0}$-sample  
    with associated hadrons for 40-80$\%$ and 0-10$\%$ $Au+Au$
    collisions}
      \end{center} 
      \label{fig:1} 
\end{figure}
Figure 1 (left) shows the azimuthal correlation for inclusive photon triggers for the most peripheral and central bins "in Au+Au collisions". 
Parton energy loss in the medium causes the away-side to be increasingly suppressed with centrality as it was previously reported
[8,16].
The suppression of the near-side yield with centrality, which has not been observed in the charged hadron 
azimuthal correlation, is consistent with an increase of the $\gamma$/$\pi^{0}$ ratio with centrality at high E$_{T}^{trig}$. 
The shower shape analysis is used to distinguish between the $(2\gamma$)/$(1\gamma)$ showers as in Figure 1 (right) 
which shows the azimuthal correlation for $\gamma$-rich sample triggers and $\pi^{0}$ triggers for the
most peripheral and central bins. The $\gamma$-rich sample has lower near-side yield than $\pi^{0}$ but not zero.
The non-zero near-side yield for the $\gamma$-rich sample is expected due to the remaining contributions of the widely 
separated photons from other
sources. The shower shape analysis is only effective for the two close $\gamma$ showers. 

The purity of using the shower shape analysis in $\pi^{0}$
identification is verified by comparing to previous measurements of azimuthal correlations between charged hadrons ($ch-ch$) [16]. 
\begin{figure}
\begin{center}
\begin{tabular}{cc}
   \resizebox{120mm}{!}{\includegraphics{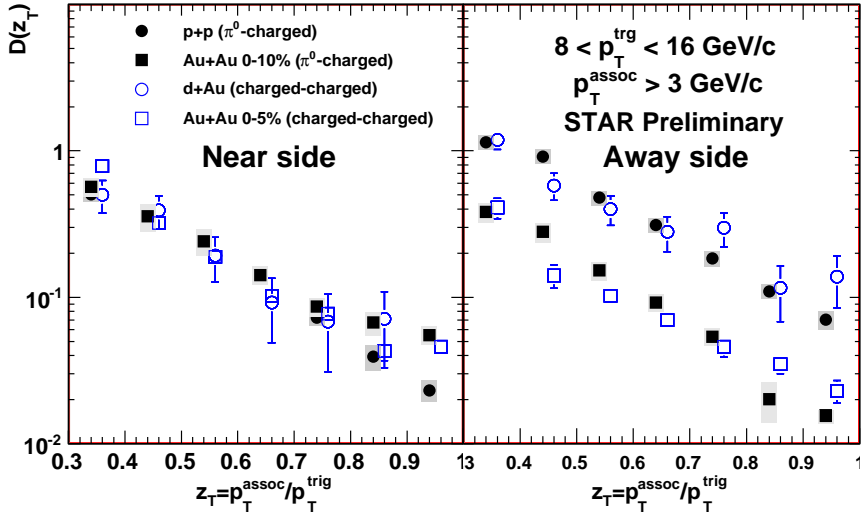}} 
        \end{tabular}
    \caption{$z_{T}$ dependence of $\pi^{0}$-ch and $ch-ch$ [16] near-side (left panel) and away-side (right panel) associated particle yields.}
      \end{center} 
      \label{fig:2} 
\end{figure}
Figure 2 shows the $z_{T}$ dependence of the associated hadron yield normalized per $\pi^{0}$ trigger D($z_{T}$), where $z_{T}= p_{T}^{assoc}/p_{T}^{trig}$
[17], for the near-side and away-side compared to
the per charged hadron trigger [16]. The near-side yield as in
Figure 2 (left) shows no significant difference between $p+p$, $d+Au$, and $Au+Au$ indicating in-vacuum fragmentation even in 
heavy ion collisions, This can be due to either a
surface bias as generated in several model calculations [18,19,20,21] or the parton fragmenting in vacuum after losing energy in the
medium. However the medium effect is clearly 
seen in the away-side in Figure 2 (right) where the per trigger yield in
$Au+Au$ is significantly suppressed compared to $p+p$ and $d+Au$. The general agreement between 
the results from this analysis
($\pi^{0}$-$ch$) and the previous analysis ($ch-ch$) is clearly seen in both panels of Figure 2 
which indicates the
purity of the $\pi^{0}$ sample and therefore the effectiveness of shower shape analysis in $\pi^{0}$ identifications.

The away-side associated yields per trigger photon for direct $\gamma$-charged hadron correlations are extracted using Eq. 1.  
Figure 3 (left) shows the $z_{T}$ dependence of the trigger-normalized fragmentation function for $\pi^{0}$-charged correlations 
($\pi^{0}$-ch)
compared to measurements with direct $\gamma$-charged correlations ($\gamma$-$ch$). The away-side yield per 
trigger of direct 
$\gamma$ is smaller than with $\pi^{0}$ trigger at the same centrality class. This difference is due to 
the fact that the 
$\pi^{0}$ originates from higher initial parton energy and therefore supports the energy loss at the parton level before
fragmentation as it was previously reported [22]. 
\begin{figure}
 \begin{center}
  \begin{tabular}{cc}
   \resizebox{90mm}{!}{\includegraphics{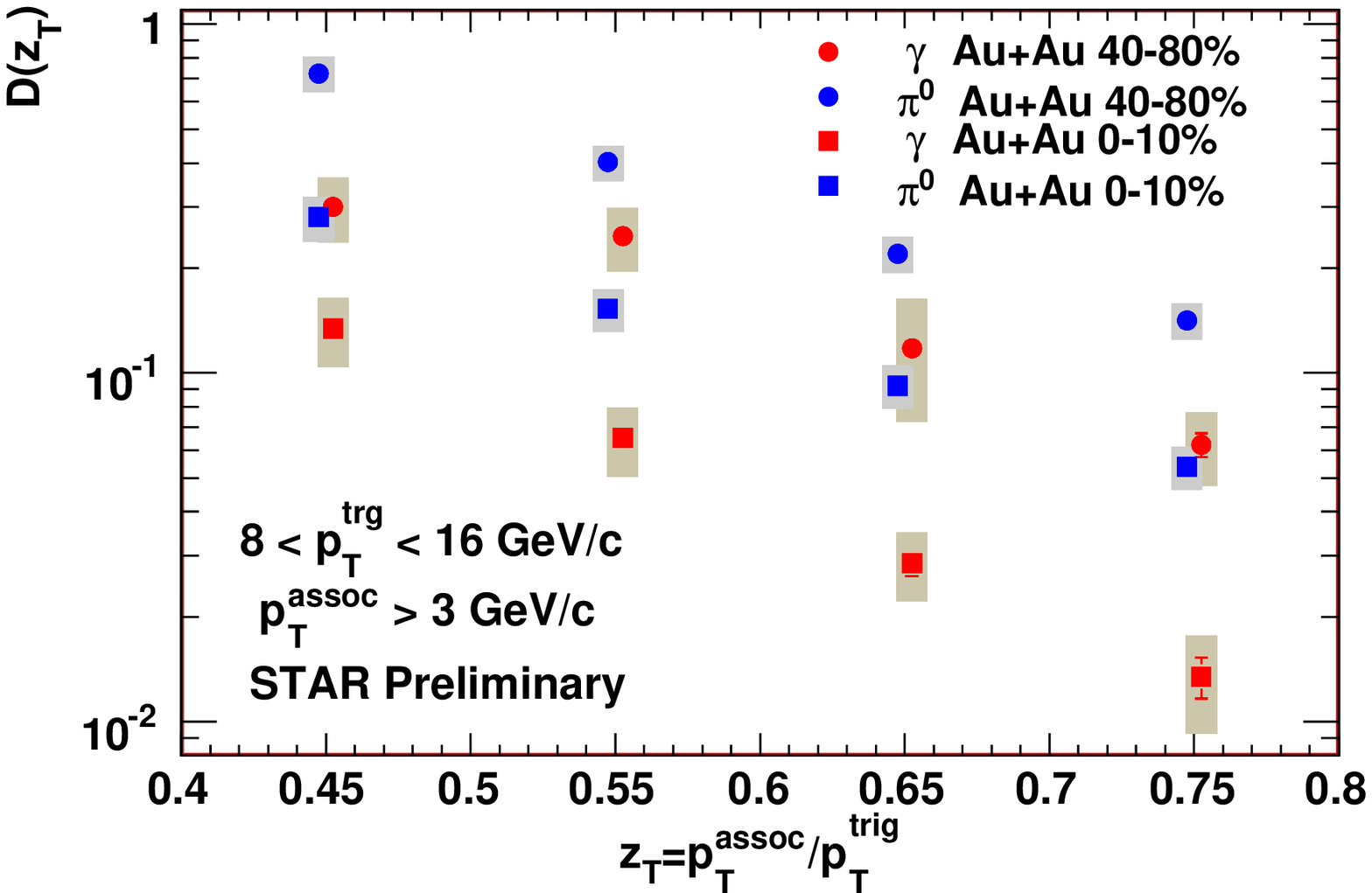}} 
   \resizebox{90mm}{!}{\includegraphics{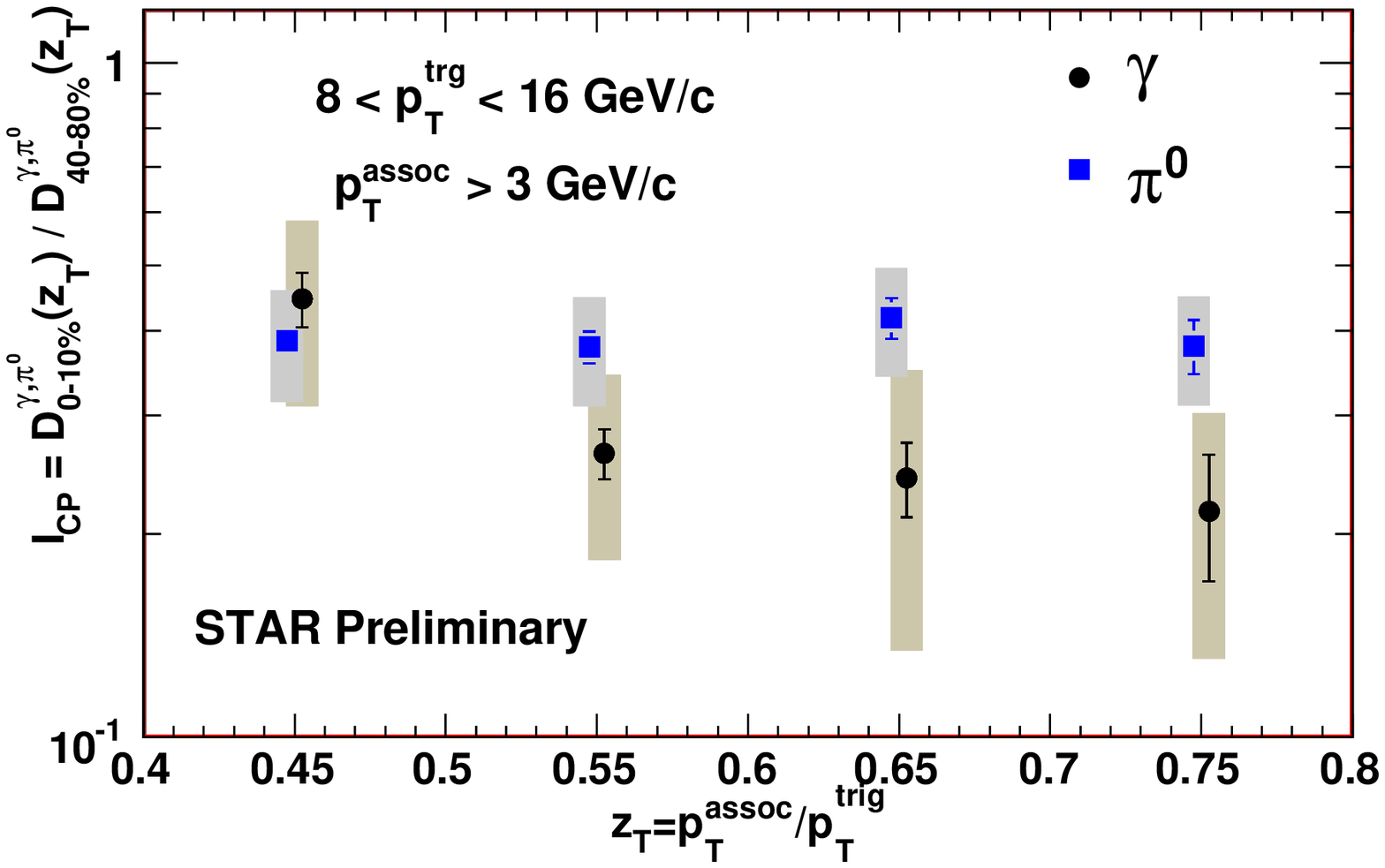}} \\
     \end{tabular}
    \caption{(Left) $z_{T}$ dependence of $\pi^{0}$ triggers and direct $\gamma$ triggers associated particle yields for 40-80$\%$ and
    0-10$\%$ $Au+Au$ collisions. (Right) $z_{T}$ dependence of I$_{CP}$ for direct $\gamma$ triggers and $\pi^{0}$ triggers (see text). Boxes
    show the systematic uncertainties.}
    \end{center}
    \label{fig:3}
\end{figure}
\begin{figure}
 \begin{center}
  \begin{tabular}{cc}
   \resizebox{90mm}{!}{\includegraphics{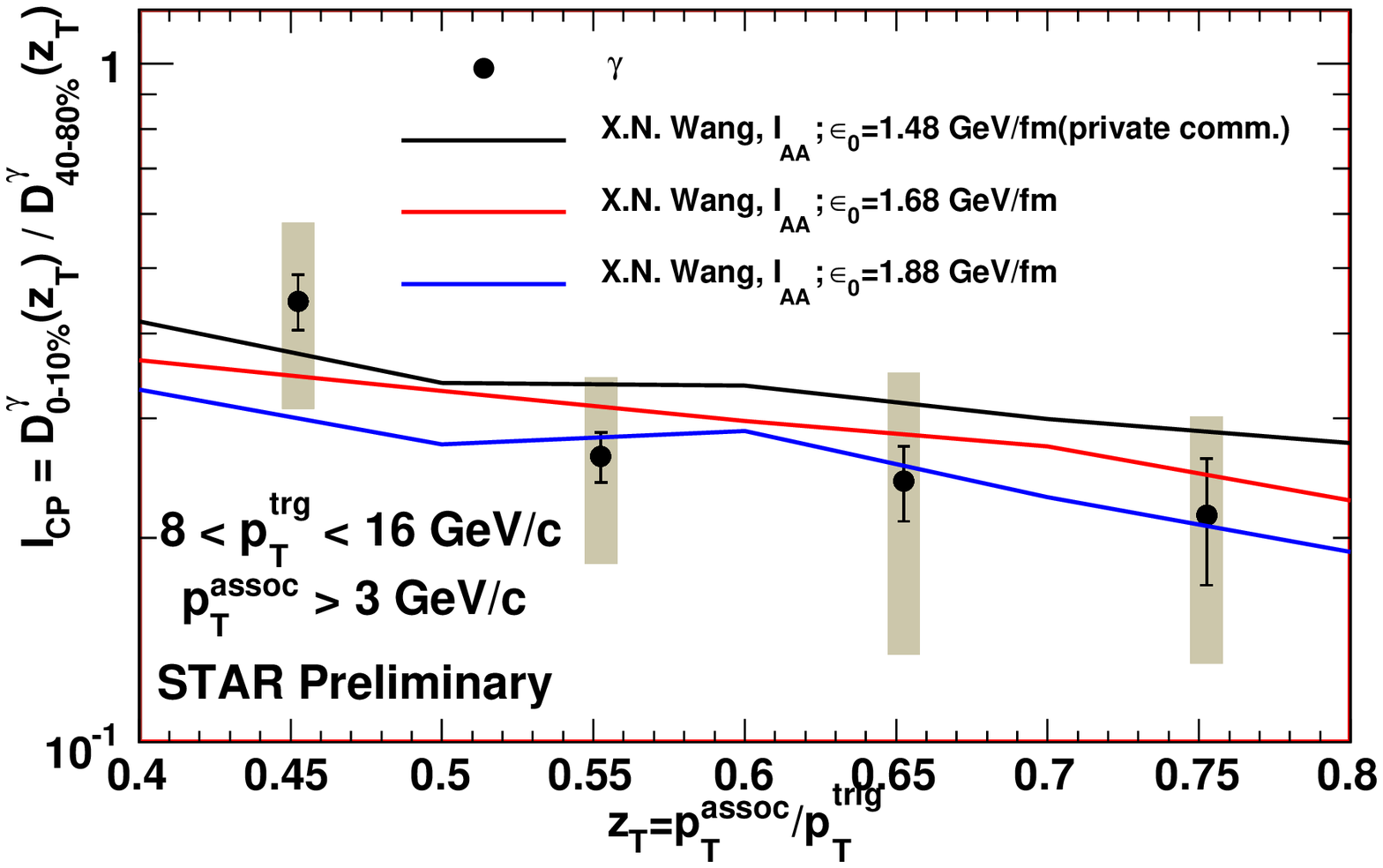}} 
   \resizebox{90mm}{!}{\includegraphics{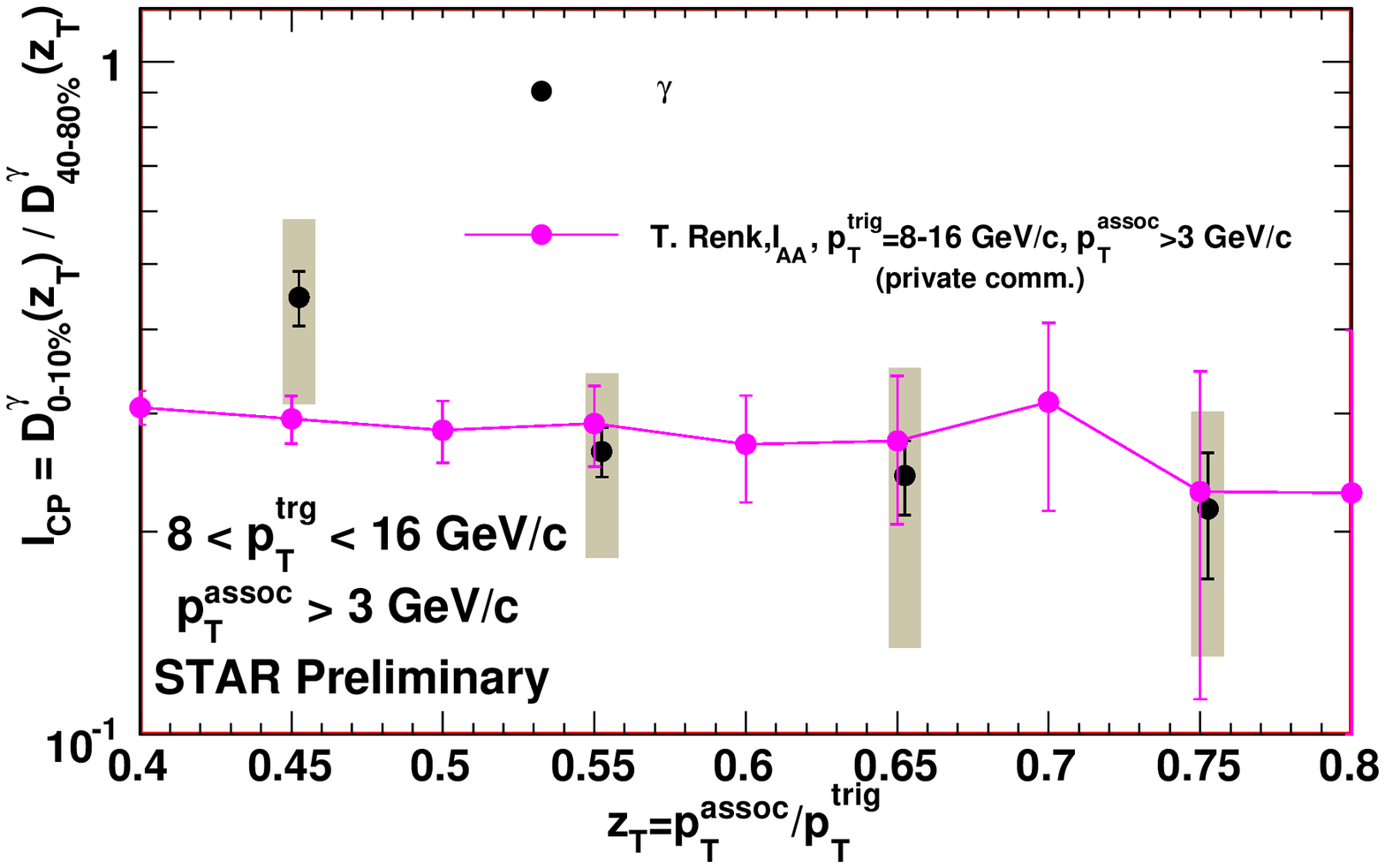}} \\
     \end{tabular}
    \caption{$z_{T}$ dependence of $I_{CP}$ for direct $\gamma$ triggers associated particle yields compared with theoretical calculations
    (left) $I_{AA}$ of 0-10$\%$ $Au+Au$ collisions (Only Annihilation and Compton processes to NLO are 
    considered) with three different initial gluon density where 7 $<$ p$_T^{trig}$ $<$ 9 GeV and p$_T^{assoc}$ $>$ 5 GeV/c, (Right) $I_{AA}$ where 
    8 $<$ p$_T^{trig}$ $<$ 16 GeV and p$_T^{assoc}$ $>$ 3 GeV/c.}
    \end{center}
    \label{fig:4}
\end{figure}

In order to quantify the away-side suppression, we calculate the quantity I$_{CP}$, which is defined as the 
ratio of the integrated yield of 
the away-side associated particles per trigger particle in $Au+Au$ central 0-10$\%$ of the geometrical cross section; 
relative to $Au+Au$
peripheral 40-80$\%$ of the geometrical cross section collisions.
Figure 3 (right) shows the I$_{CP}$ for $\pi^{0}$ triggers and for direct $\gamma$ triggers 
as a function of $z_{T}$. The ratio would be unity if there were no medium effects on the parton fragmentation; 
indeed the ratio deviates
from unity by a factor of $\sim$ 2.5. The ratio for the $\pi^{0}$ trigger is approximately independent of $z_{T}$ for the shown
range in agreement with the previous results from ($ch-ch$) measurements [16]. 
Within the current systematic uncertainty the I$_{CP}$ of direct $\gamma$ and $\pi^{0}$ are similar.    

Suppression ratios with respect to the p+p reference, I$_{AA}$, have
been reported earlier [23]. The values of I$_{AA}$ are smaller than for I$_{CP}$,
indicating finite suppression in the peripheral 40-80$\%$ data, but the
statistical uncertainties are large due to the small $\gamma$/$\pi^{0}$ ratio in p+p as previously reported [24]. Although 
the value of I$_{AA}$ is found to be similar to the values observed
for di-hadron correlations and for single-particle suppression R$_{AA}$.  

A comparison of I$_{CP}$ of direct $\gamma$ with two theoretical model calculations of I$_{AA}$ of direct $\gamma$ is shown in Figure 4.
The I$_{CP}$ values agree well with the theoretical predictions within the current uncertainties.
Figure 4 (left) indicates the need for more reduction in the systematic and statistical uncertainties in order 
to distinguish between different color charge densities.   

\section{Summary and Outlook}
\label{summary and outlook}
In summary, a first measurement of fragment distributions for jets with a controlled energy via $\gamma$-jet in $Au+Au$ collisions has been performed by the 
STAR experiment. The STAR detector is unique to perform such correlation measurements due to the full coverage in azimuth. 
Within the current uncertainty the recoil suppression ratio I$_{CP}$ of direct $\gamma$ and $\pi^{0}$ are similar.
A full analysis of the systematic uncertainties is under way and may
lead to a reduction of the total uncertainty. Future RHIC runs will
provide larger data samples to further reduce the uncertainties and
extend the $z_{T}$ range.

\end{document}